\documentclass[twocolumn]{aastex7}

\usepackage{graphicx}
\usepackage{amsmath,bm}
\usepackage[normalem]{ulem}
\usepackage{url}
\usepackage{multirow}
\usepackage{appendix}
\usepackage{comment}
\usepackage[version=4]{mhchem}
\usepackage{physics}

\graphicspath{{./}{figures/}}
\shorttitle{Acceleration of Heavy Ions}
\shortauthors{Caprioli et al.}

\begin{document}

\title{Acceleration of Heavy Ions at Non-Relativistic Collisionless Shocks}

\author[0000-0003-0939-8775]{Damiano Caprioli}
\email{caprioli@uchicago.edu}
\affiliation{Department of Astronomy and Astrophysics, The University of Chicago, 5640 S Ellis Ave, Chicago, IL 60637, USA}
\affiliation{Enrico Fermi Institute, The University of Chicago, 5640 S Ellis Ave, Chicago, IL 60637, USA}

\author[0000-0002-1879-457X]{Luca Orusa}
\affiliation{Department of Astrophysical Sciences, Princeton University, Princeton, NJ 08544, USA}
\email[show]{luca.orusa@princeton.edu}  
\affiliation{Department of Astronomy and Columbia Astrophysics Laboratory, Columbia University, New York, NY 10027, USA}

\author[0000-0002-5088-1745]{Miha Cernetic}
\email{cernetic@uchicago.edu}
\affiliation{Department of Astronomy and Astrophysics, The University of Chicago, 5640 S Ellis Ave, Chicago, IL 60637, USA}
\affiliation{Enrico Fermi Institute, The University of Chicago, 5640 S Ellis Ave, Chicago, IL 60637, USA}

\author[0000-0002-2160-7288]{Colby C. Haggerty}
\email{colbyh@hawaii.edu}
\affiliation{Institute for Astronomy, University of Hawaii, 2680 Woodlawn Drive, Honolulu, HI 96822, USA}

\author[0000-0003-4912-0161]{Bricker Ostler}
\email{bostler@uchicago.edu}
\affiliation{Department of Physics, The University of Chicago, 5720 S Ellis Ave, Chicago, IL 60637, USA}

\begin{abstract}
We investigate the process of Diffusive Shock Acceleration (DSA) of particles with mass number to charge number ratios $A/Q > 1$, e.g., partially-ionized heavy ions. 
To this end, we introduce helium- and carbon-like ions at solar abundances into two-dimensional hybrid (kinetic ions--fluid electrons) simulations of non-relativistic collisionless shocks.
This study yields three main results:
1) Heavy ions are preferentially accelerated compared to hydrogen. For typical solar abundances, the energy transferred to accelerated helium ions is comparable to, or even exceeds, that of hydrogen, thereby enhancing the overall shock acceleration efficiency.
2) Accelerated helium ions contribute to magnetic field amplification, which increases the maximum attainable particle energy and steepen the spectra of accelerated particles.
3) The efficient acceleration of helium significantly enhances the production of hadronic gamma rays and neutrinos, likely dominating the one due to hydrogen.
These effects should be taken into account, especially when modeling strong space and astrophysical shocks.
\end{abstract}

\section{Introduction} \label{sec:intro}

Non-relativistic collisionless shocks, such as those in supernova remnants (SNRs), are believed to be prominent sources of high energy cosmic rays (CRs) in the Galaxy. 
Blast waves accelerate particles through diffusive shock acceleration (DSA) \citep{krymskii77,bell78a,blandford+78,axford+78}, a process that relies on repeated scattering of particles across the shock:
particle injection, maximum achievable energy, and even CR spectra depend on the strength and inclination of the background magnetic field, as well as the ability of the shock to self-generate magnetic turbulence.

Over the past few years, collisionless shocks have been studied extensively with the help of self-consistent hybrid simulations, which typically, have ionized H as the single ion species and electrons as a neutralizing fluid \citep[e.g.,][]{lipatov02}. 
After some pioneering works including $\alpha$ particles \citep{burgess89, trattner+94}, \cite{caprioli+17}, hereafter CYS17, performed large 2D simulations with ions heavier than H (hereafter referred to as \textit{heavy ions}) at such low abundances so that they act as test particles, i.e., unable to affect the evolution of the shock. 
They found that the ions  with mass number $A$ and charge number $Q$ are injected into DSA proportional to $(A/Q)^2$ for strong shocks with Mach number $M \gtrsim 10$;
such an injection enhancement is only proportional to $(A/Q)$ for weaker shocks with $M\lesssim 10$, and in general correlates with the amount of magnetic turbulence generated by the shock.
More recently, \cite{hanusch+19a} considered 1D simulations with solar abundances for heavy ions and reported an injection boost proportional to $A/Q$.
Such a difference is likely due to the reduced dimensionality of the simulations: the very reason why heavy ions are preferentially injected over H is that they can be effectively isotropized immediately downstream of the shock in the presence of strong H-induced magnetic turbulence, eventually leaking back upstream (CYS17). 
In 1D, such turbulence cannot fully develop because the component of $\vb{B}$ normal to the shock is artificially constrained to be constant. Other 1D simulations with heavier particles were also performed in \cite{schreiner+20,fang+22}, with a primary focus on the temperature ratios of the different species rather than on the injection efficiency.

Regardless of the exact scaling with $A/Q$, kinetic simulations agree that heavier ions are preferentially injected into DSA. 
With the fiducial scaling of CYS17, singly-ionized He with $A/Q=4$ has to be injected $\sim 16$ times more effectively than H. 
The solar/Galactic He/H ratio is about $10\%$, which suggests that the normalization of He and H spectra should be comparable and He should be dynamically important.

In this paper, we perform the first 2D hybrid simulations of non-relativistic collisionless shocks with heavy ions at solar abundances treated as kinetic particles.  We find that including them has a profound impact on the dynamics of the simulation and the expected observables.

\section{Simulations} \label{sec:sim}
We use the hybrid code \textit{dHybridR} \citep{haggerty+19a,gargate+07} to perform 2D-3V simulations of non-relativistic shocks. 
The code treats an arbitrary number of ion species kinetically and electrons as a massless, charge neutralizing fluid;
it also retains the fully-relativistic dynamics of the small fraction of accelerated particles. 
All species are initialized with the same temperature and electrons are evolved using an adiabatic equation of state with $\gamma_e = 5/3$. 

Following CYS17, we normalize physical quantities as follows. Time is measured in units of inverse ion cyclotron frequency $\omega_c^{-1} = mc / eB_0$, where $c$ is the speed of light, $m$ and $e$ are the proton mass and charge, and $B_0$ is the initial magnetic field strength; velocity is measured in units of the Alfv\'en speed $v_{A} = B_0 / \sqrt{4\pi m n_0}$, where $n_0$ is a reference proton number density; and length is given in units of ion skin depth $d_i = c/\omega_p$, where $\omega_p = \sqrt{4\pi n_0 e^2 / m}$ is the ion plasma frequency. 

In each simulation, the non-relativistic shock is created by sending a supersonic flow in the $-\vu{x}$ direction against a reflecting wall stationed at the left boundary. Once the initial stream and reflected stream meet, a sharp discontinuity is formed. 
The initial background magnetic field, $\vb{B}_0 = B_0 \vu{x}$, is parallel to the shock normal. 
Quasi-parallel shocks are known to efficiently inject thermal particles and strongly amplify the initial magnetic field, thus producing extended  power-law spectra \citep{caprioli+14a,caprioli+14b,caprioli+14c,caprioli+15,haggerty+20,haggerty+22}. 
The box length ranges from 10000 $d_i$ to 60000 $d_i$, depending on the simulation, sufficiently large to prevent energetic particles from escaping through the rightmost open boundary. This ensures that the effects of DSA are retained. Transverse boundary conditions are taken to be periodic with a box size of 200 $d_i$.

We run simulations with Alfvénic Mach numbers of $M=\{5,20,40\}$. 
The Mach number, $M = M_A \simeq M_s$, is defined as the ratio of the upstream bulk flow speed $v_{sh}$ (measured in the downstream rest frame) to the Alfvén speed $v_A$ (for the Alfvénic Mach number $M_A$) or to the sound speed (for the sonic Mach number $M_s$). $c$ is chosen for the different simulations such that $v_{sh}=0.1c$. We also define the quantity $E_{sh}= m v_{sh}^2/2$ to which energies are normalized.
In all simulations, we use 25 particles per cell for H and 4 particles per cell for heavier species, with a resolution of 2 cells per $d_i$.
The shock evolution is followed for hundreds of ion cyclotron times until the expected power law distribution is established.

For each $M$, we perform two simulations: one with ionized H only, and one in which we add, on top, singly-ionized He and CNO-like species with $Q=1$ and mass numbers of $A=4,\, 14$ respectively.
The relative number abundances $\chi_s = N_s / N_H$ are set to solar values, with $\chi_{\rm He} = 0.0964$ and $\chi_{\rm CNO} = 9.53 \times 10^{-4}$ \citep{lodders03}. 
This implies that the shock velocity is the same in units of the Alfvén speed measured in the H density, but with heavier elements the Mach numbers are $\sim 20\%$ larger, since both the total Alfvén and sound speeds are reduced by a factor of $\sqrt{1 + m_{\rm He}\chi_{\rm He} + m_{\rm CNO}\chi_{\rm CNO}} \sim 1.19$. 
This minor difference does not affect our results, as our focus is on the major variations across different $M$ regimes.
The next most abundant species \citep[Fe-like and Mg-like, e.g.,][]{hoerandel+06}, are expected to be dynamically negligible \citep{caprioli+11, caprioli+17}.

\section{High-$M$ regime} \label{sec:ener}

\begin{figure}[t]
\begin{center}
\includegraphics[width=0.48\textwidth, clip=true,trim= 0 690 0 0]
{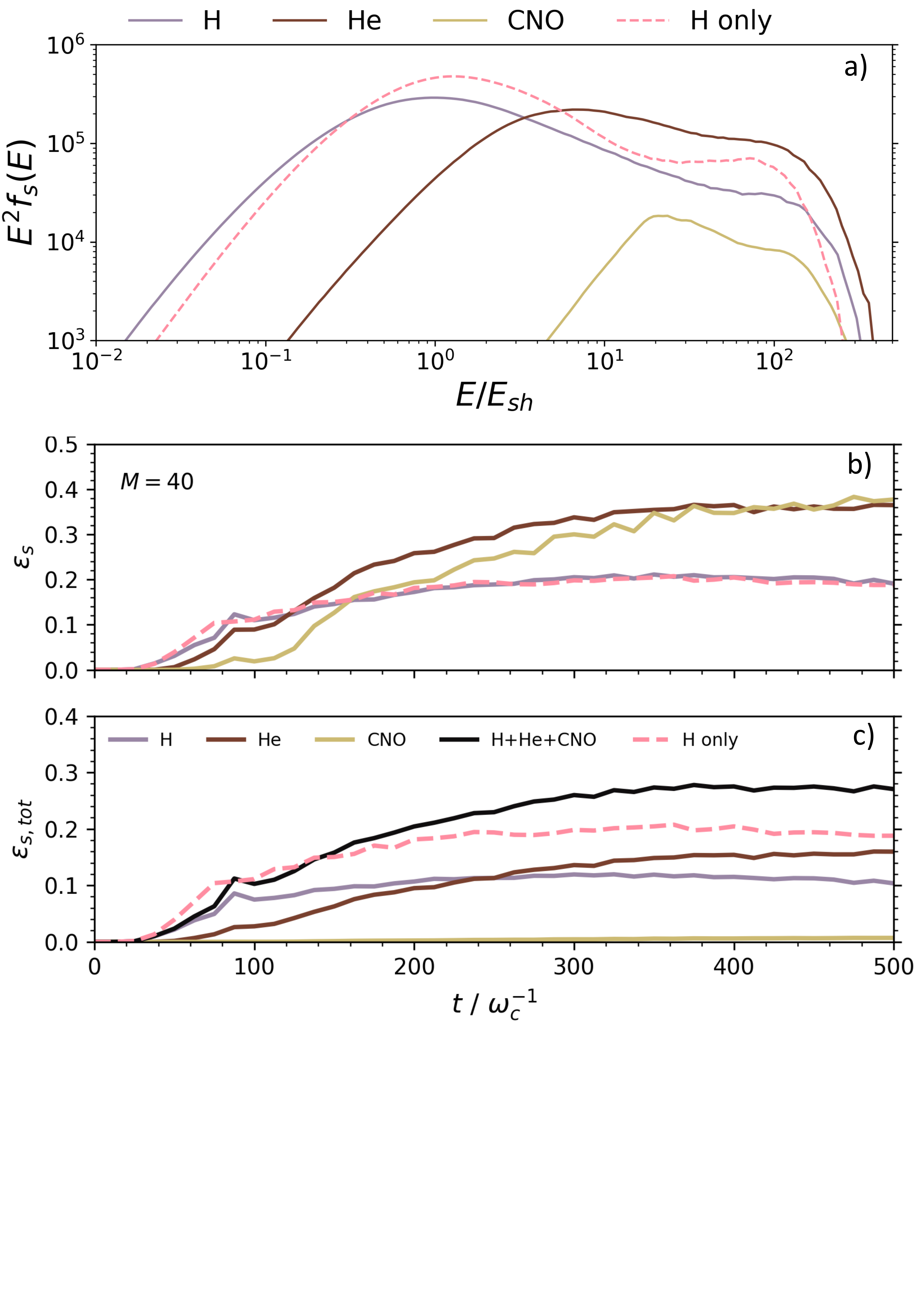}
\caption{Panel a: Downstream energy spectra from $M=40$ shock simulations with and without heavy ions at time $t=500\,\omega_c^{-1}$. Panels b and c: Injection efficiency $\varepsilon_{s}$ and total injection efficiency $\varepsilon_{s, \rm tot}$ as a function of time\label{fig:energy_compare}.}
\end{center}
\end{figure}

We first consider a benchmark run representative of strong shocks with $M = 40$. Figure~\ref{fig:energy_compare}a shows the downstream energy spectra for different species at $t = 500\,\omega_c^{-1}$, measured over an integration window of $450\, d_i$ downstream of the shock. 
This choice ensures that the spectra reflect only the latest acceleration stages, without contamination from earlier times, when injection of heavy ions has not started, yet. 
The results are compared with a run using the same parameters but including only H (dashed lines).
When heavy ions are included, the normalization of the H non-thermal tail decreases and ends up below the He tail, which dominates energetically;
the CNO ions remain subdominant, with spectra about an order of magnitude below both H and He, according to the expected enhancement $\propto (A/Q)^2$. 
In addition, H and He particles reach maximum energies $\sim 2$ times higher than H in the H-only run. 
The modification of the H spectrum and the higher maximum energies attained indicate that heavy ions play a dynamically important role in the acceleration process, as we will discuss in the next Section.
Finally, in the presence of heavy elements, the peaks of the He and CNO Maxwellians scale with $A$ as reported by CSY17, whereas the H peak does not, a result of the extra energy that the shock channels into heavy ions, which is subtracted from the heating of thermal H.
Models of X-ray bremsstrahlung or Balmer emission that rely on inferring the H temperature from Rankine--Hugoniot conditions or assume that all the species are in thermal equilibrium may be affected by this effect.

To quantify particle acceleration, for each species $s$ we compute two quantities: the \textit{individual} efficiency $\varepsilon_{s}$ and the \textit{total} efficiency $\varepsilon_{s,\rm tot}$, defined by
\begin{equation}
    \varepsilon_s \equiv \frac{\int_{E_{\rm inj,s}}^\infty E f_s(E) \dd{E}}{\int_0^\infty E f_s(E) \dd{E}};
    \quad
    \varepsilon_{s,\rm tot} \equiv \frac{\int_{E_{\rm inj,s}}^\infty E f_s(E) \dd{E}}{\sum_k\int_0^\infty E f_k(E) \dd{E}},
\end{equation}
where $f_s$ is the downstream energy  distribution function for a given species $s$, $E$ is the kinetic energy, and $E_{\rm inj,s}$ is the injection energy, i.e., the energy above which ions undergo DSA.
Thus, $\varepsilon_s$ is the ratio of the post-shock non-thermal kinetic energy in the species $s$ to the kinetic energy in that species, whereas the total efficiency is normalized to the kinetic energy in all species. 

To calculate the efficiencies, an injection energy, $E_{\rm inj,s}$, for each species is determined by fitting the energy spectra with a Maxwellian plus a power-law tail, following the method described in \cite{caprioli+14a}; 
here we adopt thresholds of $8 E_{\mathrm{sh}}$, 
$5 A_{\rm He} E_{\mathrm{sh}} $, and $3A_{\rm CNO} E_{\mathrm{sh}}$ for H, He, and CNO, respectively. 
The individual and total efficiencies for $M = 40$ are shown as a function of time over the full duration of the simulation in the bottom two panels of Figure~\ref{fig:energy_compare}; 
by the end of the simulation, the efficiencies have converged for H and He, while CNO still shows a slowly increasing trend.

As discussed in CSY17, H is the first species to be accelerated, followed by He and, subsequently, CNO particles. He begins to be preferentially accelerated over H around $t \sim 250\,\omega_c^{-1}$, as shown in Figure~\ref{fig:energy_compare}c. 
By the end of the simulation, $\varepsilon_H$ remains roughly unchanged whether or not heavy ions are present (Figure \ref{fig:energy_compare}b). 
However, the overall energy budget tells a different story: 
when heavy ions are included, H contributes only about 40\% of the total efficiency and He becomes dynamically dominant, accounting for nearly 60\% of the total efficiency. 
CNO ions remain energetically subdominant, at the percent level.
Overall, the total efficiency rises to $\sim$25\% in the simulation with heavy ions, appreciably higher than the 15\% obtained in the H-only case. 
This highlights a key result: the presence of heavy ions increases the fraction of shock energy converted into non-thermal particles.

We also note that individual efficiencies exceed 30\% for both He and CNO, compared to just 15\% for H, which is essentially the same in runs with and without heavy ions, which clearly demonstrates that shocks preferentially accelerate particles with higher mass-to-charge ratios. 
While injected H are reflected by the shock barrier \citep{caprioli+15}, heavy ions are scattered back upstream by post-shock magnetic irregularities. Their enhancement is therefore due to the fact that they are not influenced by the proton-regulated shock barrier and instead begin diffusing immediately (CYS17, \cite{orusa+25b}). 

\section{The scaling with $M$} \label{sec:ion}
We now examine how our results change as $M$ varies, focusing on spectral slopes, injection efficiency, magnetic field amplification, and the so-called \textit{ion enhancement}. 
The latter, introduced in CYS17, quantifies how efficiently ion species $s$ is accelerated relative to H, and is defined as the ratio of the ion to H distribution functions at a non-thermal energy $E$.
\begin{equation}
    K_{sp} \equiv \frac{f_s(E/Q_s)}{\chi_s f_p(E)}
\end{equation}
This ratio is well-defined for energies at which both species exhibit a non-thermal power-law tail (CYS17), i.e., above $E_{\rm inj,s}$ and before the cutoffs.

\begin{figure}[t]
\begin{center}
\includegraphics[width=0.49\textwidth, clip=true,trim= 120 0 10 0]{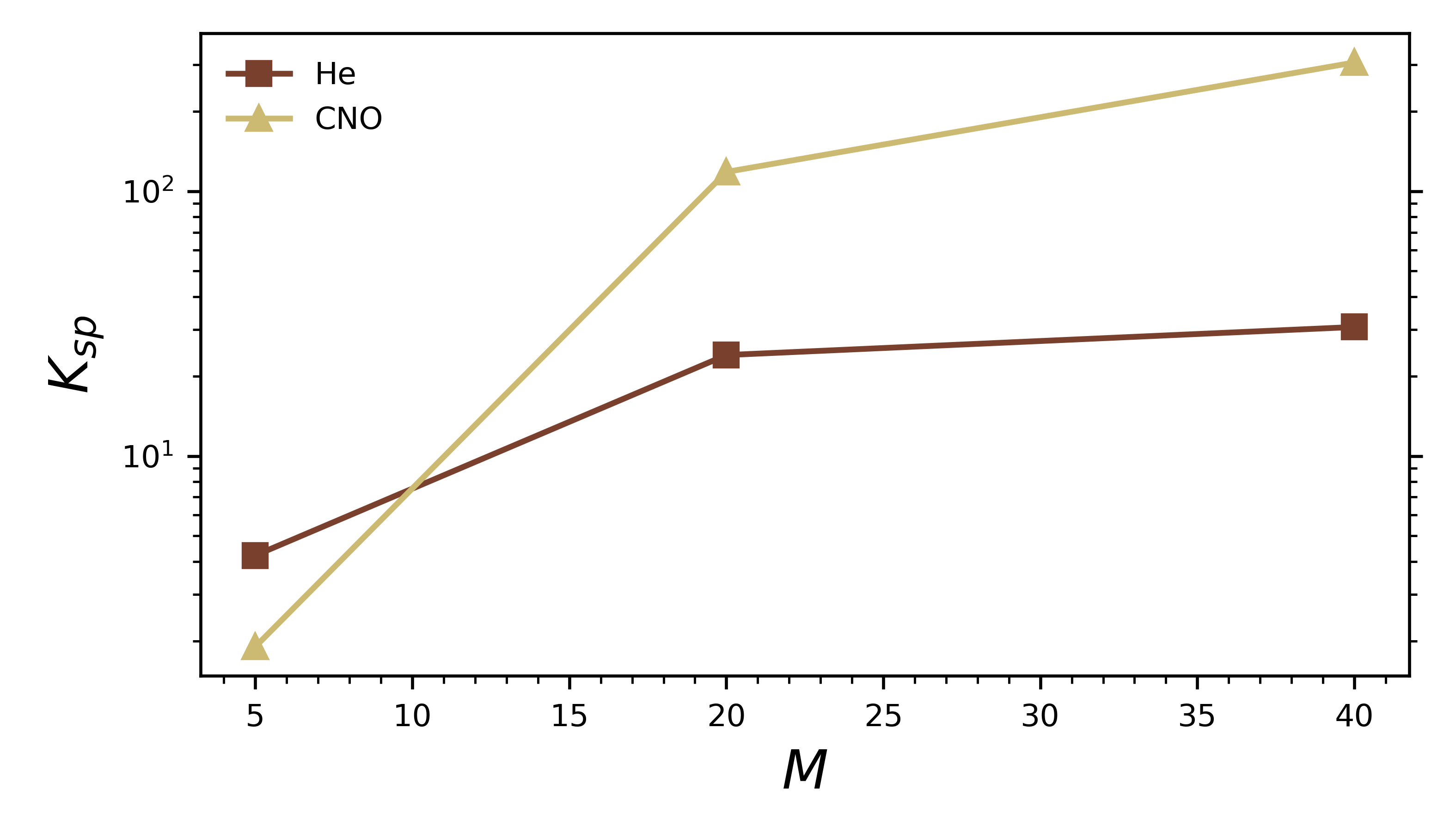}
\caption{Enhancement $K_{sp}$ of He and CNO  with respect to H as a function of $M$. 
For stronger shocks, heavy ions are preferentially injected into DSA. \label{fig:kip}}
\end{center}
\end{figure}

Figure~\ref{fig:kip} shows the ion enhancement as a function of $M$ for both He and CNO. For weaker, lower-Mach number shocks, heavy ions experience little to no enhancement in injection. However, at stronger shocks $K_{sp}$ becomes significant and increases with $M$.
For instance, at $M = 40$, He is enhanced by a factor of approximately 30, while CNO ions show an enhancement of about 250.
It is worth noting that when heavy ions are treated as test particles, the ion enhancement saturates for $M \gtrsim 10$ (see figure~2 in  CYS17). In contrast, our results show a continued increase with $M$, although the growth is less dramatic beyond $M = 20$ compared to the steep rise observed from $M = 5$ to $M = 20$. This difference, however, does not alter the $(A/Q)^2$ scaling reported in CYS17, which successfully reproduces the observed elemental abundances in Galactic CRs.
It is worth mentioning that also the sputtering of dust undergoing DSA may be a promising mechanism for heavy metal enrichment \citep[e.g.,][]{meyer+97, ellison+97, tatischeff+21}.

\begin{figure}[t]
\begin{center}
\includegraphics[width=0.48\textwidth, clip=true,trim= 80 1290 0 10]
{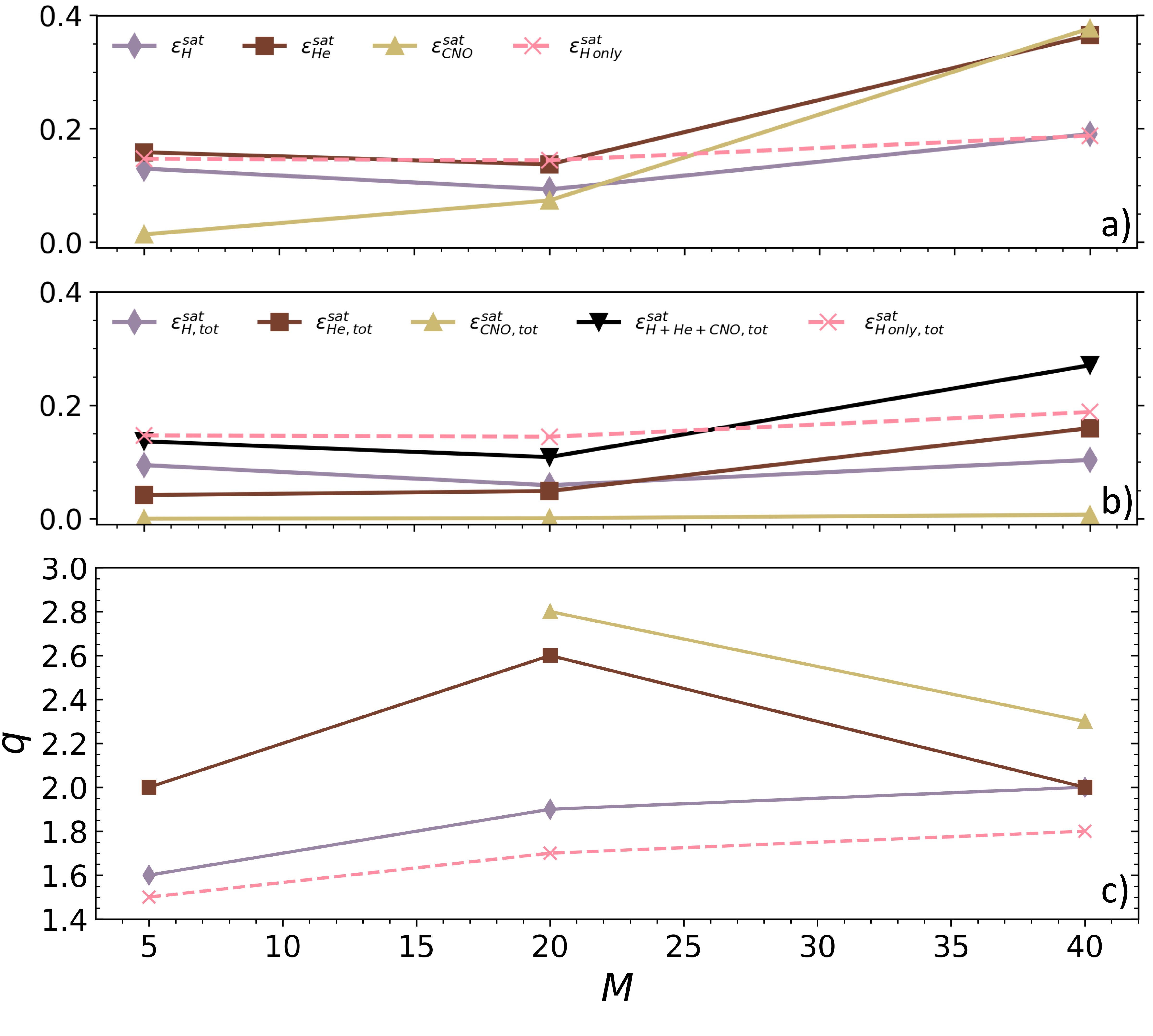}
\caption{As a function of Mach number we show the individual injection efficiency $\varepsilon_{s}$ (panel a), the total injection efficiency $\varepsilon_{s,\mathrm{tot}}$ (panel b), and the slope of the non-thermal tail for different species (panel c) at $t = 500\,\omega_c^{-1}$ (the superscript $sat$ refers to the measurement done at this specific time, when the efficiencies have saturared).\label{fig:M_dependence}}
\end{center}
\end{figure}

By examining the efficiencies in Figure~\ref{fig:M_dependence}a,b measured at $t = 500\,\omega_c^{-1}$, we identify distinct trends. 
Figure~\ref{fig:M_dependence}a shows that $\varepsilon_H$ remains largely unchanged compared to simulations without heavy ions, while He and CNO efficiencies increase with $M$, with He becoming the dominant non-thermal species $M = 40$ and CNO, due to their low abundance, remaining dynamically negligible. 
This finding reinforces the trend previously reported by CSY17 and \citet{hanusch+19a}, which highlights a clear distinction in behavior between weak ($M \lesssim 10$) and strong ($M \gtrsim 10$) shocks.

Figure~\ref{fig:M_dependence}c illustrates the slopes of different species, measured in a region of $450 \, d_i$ immediately downstream of the shock at $500 \, \omega_c^{-1}$ and shows that the inclusion of heavy ions systematically steepens the H spectra, reaching $q \approx 2$ at high $M$. 
It is worth remembering that the canonical DSA prediction at strong shocks is that spectra converge to $p^{-4}$ in momentum, which corresponds to $E^{-1.5}$/$E^{-2}$ spectra for non-relativistic/relativistic particles, respectively.
When acceleration is efficient, though, the motion of the amplified magnetic structures in the downstream (the postcursor) tends to make spectra steeper than the test-particle DSA prediction \citep{haggerty+20, caprioli+20}.
The He spectrum is steeper than the H one for low $M$ due to imperfect injection, but at $M=40$ it converges to the same slope, $q \sim 2$.
Spectra converge to the canonical $q\sim 1.5$ at low $M$, where the postcursor effect \citep{haggerty+20} is not active, but are sizably steeper for larger $M$ where magnetic field amplification in the shock precursor is more effective; 
the later evolution of the shock can affect the measurement of the actual slopes at $M=40$, since we observe flatter spectra closer to the shock, but we can conclude that they are steeper by $\Delta q\sim 0.2-0.3$ with respect to the H-only spectra.
The measured spectral indices are consistent with the injection spectra inferred from phenomenological analyses of the primary and secondary CRs fluxes reported by AMS-02 data (see Table I in the Appendix of \cite{dimauro+23} for reference). The implications of these results are further discussed in Section~\ref{sec:obs}.

All of these trends descend from the magnetic field amplification driven by streaming CRs and observed both upstream and downstream of the shock.
Amplified magnetic fields play a crucial role in the shock dynamics and in particle acceleration \citep[][]{haggerty+20,caprioli+20}, as well as in the generation of synchrotron emission from shocked regions.
While usually assumed that this amplification is driven by H only \citep[e.g.,][]{bell04,zirakashvili+08,reville+12, ohira+09, caprioli+14b, zacharegkas+24}, our simulations suggest that including heavy ions leads to consistently larger magnetic fields. 
For $M = 5$ and $M = 20$, we find that the amplification is $\lesssim$ 1.5 times greater than in the H-only case, while at $M = 40$, the amplification becomes more pronounced, reaching a factor of 2.2.
This extra field amplification, driven by accelerated He on top of H, has a twofold effect:
on one hand, it makes spectra steeper \citep{caprioli+20}, and on the other hand, it facilitates acceleration to larger energies.
When Bohm diffusion is realized \citep{caprioli+14c}, the maximum energy scales linearly with $B$, and the reported $B$ enhancement is consistent with Figure \ref{fig:energy_compare}a, where the maximum H energy is a factor $\sim2$ larger in the presence of He.

\section{Observables} \label{sec:obs}
\subsection{High-energy $\gamma$ rays and neutrinos}
He-driven turbulence may help address the long-standing question of whether Galactic sources can accelerate CRs up to the “knee” at $\sim 2$–$3$ PeV.
Current models based on the saturation of the Bell instability \citep[e.g.,][]{bell+13,cardillo+15, cristofari+22} suggest that typical SNRs may produce a sizable amount of CR H only up to $0.1$–$0.5$ PeV; 
an extra contribution to magnetic field amplification from He goes in the right direction, though the factor of $\sim 2$ that we report here for solar abundances may not be sufficient to bridge the gap.
SN exploding in high-metallicity environments such as superbubbles may fare better thanks to currents carried by ions heavier than He. However, a definitive assessment of these effects is hindered by the uncertainties in the CR chemical composition measured in the knee region and in the intrinsic difficulties of superimposing different spectra from different environments.

The production of high-energy $\gamma$ rays and neutrinos via the decay of neutral and charged pions generated in collisions of CRs with the interstellar medium is also affected by He acceleration. 
At leading order, if H is accelerated up to an energy $E_{\rm max,H}$, heavy ions yield $\gamma$ rays and neutrinos with energies up to $E_{\gamma,\nu}\sim0.1\,E_{\rm max,H} \, Q/A$. 
Though heavy ions may be partially ionized when they are injected from a warm gas, they likely become fully stripped during the DSA process \citep{morlino09}, which means that eventually $Q/A\sim 1/2$.
This implies that, when the hadronic emission is dominated by accelerated He, as typical for solar abundances, a cutoff of the $\gamma$-ray emission at a given $E_{\gamma, max}$ would correspond to a maximum CR rigidity twice as large as if produced by H only \citep[also see figure~2 of][]{caprioli+11}.

\subsection{Overall spectra of He and heavy ions}
Galactic CR data suggest that the He spectrum, as well as those of heavier ions, is harder than that of H by about 0.1 \citep{dimauro+23, pamela11, ams15b}, a phenomenon known as \emph{discrepant hardening}.
As originally suggested by \citet{malkov+12} and \cite{hanusch+19a}, a dependence on the SNR shock speed (and hence on $M$) of the chemical enhancements may be the cause of the observational conundrum.  
Even if the instantaneous spectra had the same slopes, a spectral difference can still emerge in the time-integrated CR spectrum because at late times, when $M$ becomes sufficiently small, ions injection would be shut off while H acceleration would proceed and increase the flux at low energies, eventually making the H spectrum steeper.


\section{Conclusion} \label{sec:conc}
We have performed large 2D hybrid simulations of DSA of heavy ions 
with typical solar abundances and found that for sufficiently strong shocks $M\gtrsim 20$, He is efficiently accelerated to become a major player in the shock dynamics.
Our major findings are:
\begin{itemize}
    \item Quasi-parallel shocks tend to channel more energy in non-thermal He than H, with the total shock acceleration efficiency reaching more than $\sim 25\%$ for high-$M$ shocks, compared with $\sim 15\%$ for pure H (Figure \ref{fig:energy_compare}).
    \item The current in accelerated He ions also boosts the magnetic field amplification, which effectively increases the maximum energy obtainable from DSA, which may be important for SNRs to reach the CR knee, and strengthens the effect of the postcursor \citep{caprioli+20}, which leads to steeper CR spectra compared to the H-only case.
    \item For solar abundances, He becomes the dominant source of hadronic products such as $\gamma$-rays and neutrinos; this means that, for an observed cutoff in the $\gamma$-ray emission, the maximum CR energy is a factor of 2 larger than it would be if the emission were produced by H.
    \item CNO is typically not important because its selective injection boost does not offset its low solar abundance; 
    however, in shocks propagating through material enriched in CNO about ten times more than solar, such as molecular clouds or stellar winds, it may also contribute to the shock dynamics and emission.
\end{itemize}

One final comment is that in this work we focused on quasi-parallel shocks in the warm interstellar medium, only.
Different initial abundances and ionization states, such as in solar energetic particles \citep[e.g.,][]{tylka+05, reames+13} or in superbubbles \citep[e.g.,][]{higdon+05, tatischeff+21} may lead to different scaling of the enhancements with $A/Q$ \citep{desai+16a, lee+24}.
Moreover, though oblique and quasi-perpendicular configurations hinder the spontaneous injection of thermal particles, it was recently found that high-$M$ shocks may indeed accelerate particles at non-negligible levels \citep{orusa+23, orusa+25a, orusa+25b}.
Since this requires full 3D simulations and is thus computationally very expensive, we defer such an investigation to future works.

\begin{acknowledgments}
We would like to thank the referee for having contributed to improve the manuscript; Marwah Roussi and Elyssa Brooks for having worked on preliminary runs of shocks including heavy ions; and Luke O'C. Drury and Anatoly Spitkovsky for stimulating discussions and technical/scientific advice.
This research was partially supported by NASA grant 80NSSC18K1726, NSF grants AST-2510951 and AST-2308021 to D.C., and by NSF grant PHY-2309135 to the Kavli Institute for Theoretical Physics.

L.O. acknowledges the support of the Multimessenger Plasma Physics Center (MPPC), NSF grant PHY2206607. 
We would like to thank The University of Chicago Research Computing Center for providing the computational resources to conduct this research.
\end{acknowledgments}

\bibliography{Total}

\begin{thebibliography}{}
\expandafter\ifx\csname natexlab\endcsname\relax\def\natexlab#1{#1}\fi
\providecommand{\url}[1]{\href{#1}{#1}}
\providecommand{\dodoi}[1]{doi:~\href{http://doi.org/#1}{\nolinkurl{#1}}}
\providecommand{\doeprint}[1]{\href{http://ascl.net/#1}{\nolinkurl{http://ascl.net/#1}}}
\providecommand{\doarXiv}[1]{\href{https://arxiv.org/abs/#1}{\nolinkurl{https://arxiv.org/abs/#1}}}

\bibitem[{O. Adriani {et~al.}(2011)Adriani, Barbarino, Bazilevskaya, Bellotti,
  Boezio, Bogomolov, Bonechi, Bongi, Bonvicini, Borisov, Bottai, Bruno,
  Cafagna, Campana, Carbone, Carlson, Casolino, Castellini, Consiglio,
  De~Pascale, De~Santis, De~Simone, Di~Felice, Galper, Gillard, Grishantseva,
  Jerse, Karelin, Koldashov, Krutkov, Kvashnin, Leonov, Malakhov, Malvezzi,
  Marcelli, Mayorov, Menn, Mikhailov, Mocchiutti, Monaco, Mori, Nikonov,
  Osteria, Palma, Papini, Pearce, Picozza, Pizzolotto, Ricci, Ricciarini,
  Rossetto, Sarkar, Simon, Sparvoli, Spillantini, Stozhkov, Vacchi, Vannuccini,
  Vasilyev, Voronov, Yurkin, Wu, Zampa, Zampa, \& Zverev}]{pamela11}
Adriani, O., Barbarino, G.~C., Bazilevskaya, G.~A., {et~al.} 2011,
  \bibinfo{title}{PAMELA Measurements of Cosmic-Ray Proton and Helium Spectra,}
  Science, 332, 69, \dodoi{10.1126/science.1199172}

\bibitem[{M. Aguilar {et~al.}(2015)Aguilar, Aisa, Alpat, Alvino, Ambrosi,
  Andeen, Arruda, Attig, Azzarello, Bachlechner, Barao, Barrau, Barrin,
  Bartoloni, Basara, Battarbee, Battiston, Bazo, Becker, Behlmann, Beischer,
  Berdugo, Bertucci, Bindi, Bizzaglia, Bizzarri, Boella, de~Boer, Bollweg,
  Bonnivard, Borgia, Borsini, Boschini, Bourquin, Burger, Cadoux, Cai, Capell,
  Caroff, Casaus, Castellini, Cernuda, Cerreta, Cervelli, Chae, Chang, Chen,
  Chen, Chen, Chen, Cheng, Chou, Choumilov, Choutko, Chung, Clark, Clavero,
  Coignet, Consolandi, Contin, Corti, Gil, Coste, Creus, Crispoltoni, Cui, Dai,
  Delgado, Della~Torre, Demirk\"oz, Derome, Di~Falco, Di~Masso, Dimiccoli,
  D\'{\i}az, von Doetinchem, Donnini, Duranti, D'Urso, Egorov, Eline, Eppling,
  Eronen, Fan, Farnesini, Feng, Fiandrini, Fiasson, Finch, Fisher, Formato,
  Galaktionov, Gallucci, Garc\'{\i}a, Garc\'{\i}a-L\'opez, Gargiulo, Gast,
  Gebauer, Gervasi, Ghelfi, Giovacchini, Goglov, Gong, Goy, Grabski, Grandi,
  Graziani, Guandalini, Guerri, Guo, Haas, Habiby, Haino, Han, He, Heil,
  Hoffman, Hsieh, Huang, Huh, Incagli, Ionica, Jang, Jinchi, Kanishev, Kim,
  Kim, Kirn, Korkmaz, Kossakowski, Kounina, Kounine, Koutsenko, Krafczyk,
  La~Vacca, Laudi, Laurenti, Lazzizzera, Lebedev, Lee, Lee, Leluc, Li, Li, Li,
  Li, Li, Li, Li, Li, Li, Li, Lim, Lin, Lipari, Lippert, Liu, Liu, Liu, Lolli,
  Lomtadze, Lu, Lu, Lu, Luebelsmeyer, Luo, Luo, Lv, Majka, Ma\~n\'a,
  Mar\'{\i}n, Martin, Mart\'{\i}nez, Masi, Maurin, Menchaca-Rocha, Meng, Mo,
  Morescalchi, Mott, M\"uller, Nelson, Ni, Nikonov, Nozzoli, Nunes, Obermeier,
  Oliva, Orcinha, Palmonari, Palomares, Paniccia, Papi, Pauluzzi, Pedreschi,
  Pensotti, Pereira, Picot-Clemente, Pilo, Piluso, Pizzolotto, Plyaskin, Pohl,
  Poireau, Putze, Quadrani, Qi, Qin, Qu, R\"aih\"a, Rancoita, Rapin, Ricol,
  Rodr\'{\i}guez, Rosier-Lees, Rozhkov, Rozza, Sagdeev, Sandweiss, Saouter,
  Schael, Schmidt, von Dratzig, Schwering, Scolieri, Seo, Shan, Shan, Shi, Shi,
  Shi, Siedenburg, Son, Song, Spada, Spinella, Sun, Sun, Tacconi, Tang, Tang,
  Tang, Tao, Tescaro, Ting, Ting, Tomassetti, Torsti,
  T\"urko\ifmmode~\breve{g}\else \u{g}\fi{}lu, Urban, Vagelli, Valente,
  Vannini, Valtonen, Vaurynovich, Vecchi, Velasco, Vialle, Vitale, Vitillo,
  Wang, Wang, Wang, Wang, Wang, Wang, Weng, Whitman, Wienkenh\"over,
  Willenbrock, Wu, Wu, Xia, Xie, Xie, Xiong, Xu, Xu, Yan, Yang, Yang, Yang, Ye,
  Yi, Yu, Yu, Zeissler, Zhang, Zhang, Zhang, Zhang, Zhang, Zhang, Zhang, Zheng,
  Zhuang, Zhukov, Zichichi, Zimmermann, \& Zuccon}]{ams15b}
Aguilar, M., Aisa, D., Alpat, B., {et~al.} 2015, \bibinfo{title}{Precision
  Measurement of the Helium Flux in Primary Cosmic Rays of Rigidities 1.9 GV to
  3 TV with the Alpha Magnetic Spectrometer on the International Space
  Station,} Phys. Rev. Lett., 115, 211101,
  \dodoi{10.1103/PhysRevLett.115.211101}

\bibitem[{W.~I. {Axford} {et~al.}(1978){Axford}, {Leer}, \&
  {Skadron}}]{axford+78}
{Axford}, W.~I., {Leer}, E., \& {Skadron}, G. 1978, \bibinfo{title}{{The
  acceleration of cosmic rays by shock waves},} in International Cosmic Ray
  Conference, Vol.~11, ICRC, 132--137.
\newblock \url{http://adsabs.harvard.edu/abs/1978ICRC...11..132A}

\bibitem[{A.~R. {Bell}(1978){Bell}}]{bell78a}
{Bell}, A.~R. 1978, \bibinfo{title}{{The acceleration of cosmic rays in shock
  fronts. I},} MNRAS, 182, 147.
\newblock \url{https://ui.adsabs.harvard.edu/abs/1978MNRAS.182..147B/abstract}

\bibitem[{A.~R. Bell(2004)Bell}]{bell04}
Bell, A.~R. 2004, \bibinfo{title}{{Turbulent amplification of magnetic field
  and diffusive shock acceleration of cosmic rays},} MNRAS, 353, 550,
  \dodoi{10.1111/j.1365-2966.2004.08097.x}

\bibitem[{A.~R. {Bell} {et~al.}(2013){Bell}, {Schure}, {Reville}, \&
  {Giacinti}}]{bell+13}
{Bell}, A.~R., {Schure}, K.~M., {Reville}, B., \& {Giacinti}, G. 2013,
  \bibinfo{title}{{Cosmic-ray acceleration and escape from supernova
  remnants},} MNRAS, 431, 415, \dodoi{10.1093/mnras/stt179}

\bibitem[{R.~D. {Blandford} \& J.~P. {Ostriker}(1978){Blandford} \&
  {Ostriker}}]{blandford+78}
{Blandford}, R.~D., \& {Ostriker}, J.~P. 1978, \bibinfo{title}{{Particle
  acceleration by astrophysical shocks},} ApJL, 221, L29,
  \dodoi{10.1086/182658}

\bibitem[{D. {Burgess}(1989){Burgess}}]{burgess89}
{Burgess}, D. 1989, \bibinfo{title}{{Alpha particles in field-aligned beams
  upstream of the bow shock - Simulations},} \grl, 16, 163,
  \dodoi{10.1029/GL016i002p00163}

\bibitem[{D. {Caprioli} {et~al.}(2011){Caprioli}, {Blasi}, \&
  {Amato}}]{caprioli+11}
{Caprioli}, D., {Blasi}, P., \& {Amato}, E. 2011, \bibinfo{title}{{Non-linear
  diffusive acceleration of heavy nuclei in supernova remnant shocks},} APh,
  34, 447, \dodoi{10.1016/j.astropartphys.2010.10.011}

\bibitem[{D. {Caprioli} {et~al.}(2020){Caprioli}, {Haggerty}, \&
  {Blasi}}]{caprioli+20}
{Caprioli}, D., {Haggerty}, C.~C., \& {Blasi}, P. 2020,
  \bibinfo{title}{{Kinetic Simulations of Cosmic-Ray-modified Shocks. II.
  Particle Spectra},} \apj, 905, 2, \dodoi{10.3847/1538-4357/abbe05}

\bibitem[{D. {Caprioli} {et~al.}(2015){Caprioli}, {Pop}, \&
  {Spitkovsky}}]{caprioli+15}
{Caprioli}, D., {Pop}, A., \& {Spitkovsky}, A. 2015,
  \bibinfo{title}{{Simulations and Theory of Ion Injection at Non-relativistic
  Collisionless Shocks},} \apjl, 798, 28.
\newblock \doarXiv{1409.8291}

\bibitem[{D. {Caprioli} \& A. {Spitkovsky}(2014{\natexlab{a}}){Caprioli} \&
  {Spitkovsky}}]{caprioli+14a}
{Caprioli}, D., \& {Spitkovsky}, A. 2014{\natexlab{a}},
  \bibinfo{title}{{Simulations of Ion Acceleration at Non-relativistic Shocks:
  I. Acceleration Efficiency},} \apj, 783, 91,
  \dodoi{10.1088/0004-637X/783/2/91}

\bibitem[{D. {Caprioli} \& A. {Spitkovsky}(2014{\natexlab{b}}){Caprioli} \&
  {Spitkovsky}}]{caprioli+14b}
{Caprioli}, D., \& {Spitkovsky}, A. 2014{\natexlab{b}},
  \bibinfo{title}{{Simulations of Ion Acceleration at Non-relativistic Shocks:
  II. Magnetic Field Amplification},} \apj, 794, 46,
  \dodoi{10.1088/0004-637X/794/1/46}

\bibitem[{D. {Caprioli} \& A. {Spitkovsky}(2014{\natexlab{c}}){Caprioli} \&
  {Spitkovsky}}]{caprioli+14c}
{Caprioli}, D., \& {Spitkovsky}, A. 2014{\natexlab{c}},
  \bibinfo{title}{{Simulations of Ion Acceleration at Non-relativistic Shocks.
  III. Particle Diffusion},} \apj, 794, 47, \dodoi{10.1088/0004-637X/794/1/47}

\bibitem[{D. {Caprioli} {et~al.}(2017){Caprioli}, {Yi}, \&
  {Spitkovsky}}]{caprioli+17}
{Caprioli}, D., {Yi}, D.~T., \& {Spitkovsky}, A. 2017,
  \bibinfo{title}{{Chemical Enhancements in Shock-Accelerated Particles: Ab
  initio Simulations},} \prl, 119, 171101,
  \dodoi{10.1103/PhysRevLett.119.171101}

\bibitem[{M. Cardillo {et~al.}(2015)Cardillo, Amato, \& Blasi}]{cardillo+15}
Cardillo, M., Amato, E., \& Blasi, P. 2015, \bibinfo{title}{On the cosmic ray
  spectrum from type II supernovae expanding in their red giant presupernova
  wind,} Astroparticle Physics, 69, 1,
  \dodoi{https://doi.org/10.1016/j.astropartphys.2015.03.002}

\bibitem[{P. {Cristofari} {et~al.}(2022){Cristofari}, {Blasi}, \&
  {Caprioli}}]{cristofari+22}
{Cristofari}, P., {Blasi}, P., \& {Caprioli}, D. 2022,
  \bibinfo{title}{{Microphysics of Diffusive Shock Acceleration: Impact on the
  Spectrum of Accelerated Particles},} \apj, 930, 28,
  \dodoi{10.3847/1538-4357/ac6182}

\bibitem[{M.~I. Desai {et~al.}(2016)Desai, Mason, Dayeh, Ebert, Mccomas, Li,
  Cohen, Mewaldt, Schwadron, \& Smith}]{desai+16a}
Desai, M.~I., Mason, G.~M., Dayeh, M.~A., {et~al.} 2016,
  \bibinfo{title}{Spectral Properties of Large Gradual Solar Energetic Particle
  Events. I. Fe, O, and Seed Material,} \apj, 816, 68,
  \dodoi{10.3847/0004-637X/816/2/68}

\bibitem[{M. Di~Mauro {et~al.}(2023)Di~Mauro, Donato, Korsmeier, Manconi, \&
  Orusa}]{dimauro+23}
Di~Mauro, M., Donato, F., Korsmeier, M., Manconi, S., \& Orusa, L. 2023,
  \bibinfo{title}{Novel prediction for secondary positrons and electrons in the
  Galaxy,} Physical Review D, 108, \dodoi{10.1103/physrevd.108.063024}

\bibitem[{D.~C. {Ellison} {et~al.}(1997){Ellison}, {Drury}, \&
  {Meyer}}]{ellison+97}
{Ellison}, D.~C., {Drury}, L.~O., \& {Meyer}, J.-P. 1997,
  \bibinfo{title}{{Galactic Cosmic Rays from Supernova Remnants. II. Shock
  Acceleration of Gas and Dust},} \apj, 487, 197, \dodoi{10.1086/304580}

\bibitem[{J. Fang {et~al.}(2022)Fang, Xia, Tian, Zhou, \& Yu}]{fang+22}
Fang, J., Xia, Q., Tian, S., Zhou, L., \& Yu, H. 2022, \bibinfo{title}{Kinetic
  simulation of electron, proton and helium acceleration in a non-relativistic
  quasi-parallel shock,} Monthly Notices of the Royal Astronomical Society,
  512, 5418–5422, \dodoi{10.1093/mnras/stac886}

\bibitem[{L. {Gargat{\'e}} {et~al.}(2007){Gargat{\'e}}, {Bingham}, {Fonseca},
  \& {Silva}}]{gargate+07}
{Gargat{\'e}}, L., {Bingham}, R., {Fonseca}, R.~A., \& {Silva}, L.~O. 2007,
  \bibinfo{title}{{dHybrid: A massively parallel code for hybrid simulations of
  space plasmas},} Computer Physics Communications, 176, 419,
  \dodoi{10.1016/j.cpc.2006.11.013}

\bibitem[{C.~C. {Haggerty} {et~al.}(2022){Haggerty}, {Bret}, \&
  {Caprioli}}]{haggerty+22}
{Haggerty}, C.~C., {Bret}, A., \& {Caprioli}, D. 2022, \bibinfo{title}{{Kinetic
  simulations of strongly magnetized parallel shocks: deviations from MHD jump
  conditions},} \mnras, 509, 2084, \dodoi{10.1093/mnras/stab3110}

\bibitem[{C.~C. {Haggerty} \& D. {Caprioli}(2019){Haggerty} \&
  {Caprioli}}]{haggerty+19a}
{Haggerty}, C.~C., \& {Caprioli}, D. 2019, \bibinfo{title}{{dHybridR: A Hybrid
  Particle-in-cell Code Including Relativistic Ion Dynamics},} \apj, 887, 165,
  \dodoi{10.3847/1538-4357/ab58c8}

\bibitem[{C.~C. {Haggerty} \& D. {Caprioli}(2020){Haggerty} \&
  {Caprioli}}]{haggerty+20}
{Haggerty}, C.~C., \& {Caprioli}, D. 2020, \bibinfo{title}{{Kinetic Simulations
  of Cosmic-Ray-modified Shocks. I. Hydrodynamics},} \apj, 905, 1,
  \dodoi{10.3847/1538-4357/abbe06}

\bibitem[{A. {Hanusch} {et~al.}(2019){Hanusch}, {Liseykina}, \&
  {Malkov}}]{hanusch+19a}
{Hanusch}, A., {Liseykina}, T.~V., \& {Malkov}, M. 2019,
  \bibinfo{title}{{Acceleration of Cosmic Rays in Supernova Shocks: Elemental
  Selectivity of the Injection Mechanism},} \apj, 872, 108,
  \dodoi{10.3847/1538-4357/aafdae}

\bibitem[{J.~C. {Higdon} \& R.~E. {Lingenfelter}(2005){Higdon} \&
  {Lingenfelter}}]{higdon+05}
{Higdon}, J.~C., \& {Lingenfelter}, R.~E. 2005, \bibinfo{title}{{OB
  Associations, Supernova-generated Superbubbles, and the Source of Cosmic
  Rays},} ApJ, 628, 738, \dodoi{10.1086/430814}

\bibitem[{J.~R. {H{\"o}randel et al.}(2006){H{\"o}randel et
  al.}}]{hoerandel+06}
{H{\"o}randel et al.}, J.~R. 2006, \bibinfo{title}{{Results from the KASCADE,
  KASCADE-Grande, and LOPES experiments},} Journal of Physics Conference
  Series, 39, 463, \dodoi{10.1088/1742-6596/39/1/122}

\bibitem[{G.~F. {Krymskii}(1977){Krymskii}}]{krymskii77}
{Krymskii}, G.~F. 1977, \bibinfo{title}{{A regular mechanism for the
  acceleration of charged particles on the front of a shock wave},} Akademiia
  Nauk SSSR Doklady, 234, 1306.
\newblock \url{https://ui.adsabs.harvard.edu/abs/1977DoSSR.234R1306K}

\bibitem[{J.-Y. {Lee} {et~al.}(2024){Lee}, {Kahler}, {Raymond}, \&
  {Ko}}]{lee+24}
{Lee}, J.-Y., {Kahler}, S., {Raymond}, J.~C., \& {Ko}, Y.-K. 2024,
  \bibinfo{title}{{Solar Energetic Particle Charge States and Abundances with
  Nonthermal Electrons},} \apj, 963, 70, \dodoi{10.3847/1538-4357/ad1ab6}

\bibitem[{A.~S. {Lipatov}(2002){Lipatov}}]{lipatov02}
{Lipatov}, A.~S. 2002, {\emph{The hybrid multiscale simulation technology: an
  introduction with application to astrophysical and laboratory plasmas}}
  (Berlin; New York: Springer,~Scientific computation)

\bibitem[{K. {Lodders}(2003){Lodders}}]{lodders03}
{Lodders}, K. 2003, \bibinfo{title}{{Solar System Abundances and Condensation
  Temperatures of the Elements},} \apj, 591, 1220, \dodoi{10.1086/375492}

\bibitem[{M.~A. Malkov {et~al.}(2012)Malkov, Diamond, \& Sagdeev}]{malkov+12}
Malkov, M.~A., Diamond, P.~H., \& Sagdeev, R.~Z. 2012,
  \bibinfo{title}{Proton-Helium Spectral Anomaly as a Signature of Cosmic Ray
  Accelerator,} Physical Review Letters, 108, 081104,
  \dodoi{10.1103/PhysRevLett.108.081104}

\bibitem[{J. {Meyer} {et~al.}(1997){Meyer}, {Drury}, \& {Ellison}}]{meyer+97}
{Meyer}, J., {Drury}, L.~O., \& {Ellison}, D.~C. 1997,
  \bibinfo{title}{{Galactic Cosmic Rays from Supernova Remnants. I. A
  Cosmic-Ray Composition Controlled by Volatility and Mass-to-Charge Ratio},}
  ApJ, 487, 182, \dodoi{10.1086/304599}

\bibitem[{G. {Morlino}(2009){Morlino}}]{morlino09}
{Morlino}, G. 2009, \bibinfo{title}{{Cosmic-Ray Electron Injection from the
  Ionization of Nuclei},} Physical Review Letters, 103, 121102,
  \dodoi{10.1103/PhysRevLett.103.121102}

\bibitem[{Y. {Ohira} {et~al.}(2009){Ohira}, {Reville}, {Kirk}, \&
  {Takahara}}]{ohira+09}
{Ohira}, Y., {Reville}, B., {Kirk}, J.~G., \& {Takahara}, F. 2009,
  \bibinfo{title}{{Two-Dimensional Particle-In-Cell Simulations of the
  Nonresonant, Cosmic-Ray-Driven Instability in Supernova Remnant Shocks},}
  \apj, 698, 445, \dodoi{10.1088/0004-637X/698/1/445}

\bibitem[{L. {Orusa} \& D. {Caprioli}(2023){Orusa} \& {Caprioli}}]{orusa+23}
{Orusa}, L., \& {Caprioli}, D. 2023, \bibinfo{title}{{Fast Particle
  Acceleration in 3D Hybrid Simulations of Quasiperpendicular Shocks},} \prl,
  131, 095201, \dodoi{10.1103/PhysRevLett.131.095201}

\bibitem[{L. Orusa {et~al.}(2025)Orusa, Caprioli, Sironi, \&
  Spitkovsky}]{orusa+25b}
Orusa, L., Caprioli, D., Sironi, L., \& Spitkovsky, A. 2025, The role of
  three-dimensional effects on ion injection and acceleration in perpendicular
  shocks, \doarXiv{2507.13436}

\bibitem[{L. Orusa \& V. Valenzuela-Villaseca(2025)Orusa \&
  Valenzuela-Villaseca}]{orusa+25a}
Orusa, L., \& Valenzuela-Villaseca, V. 2025, \bibinfo{title}{Criteria for ion
  acceleration in laboratory magnetized quasi-perpendicular collisionless
  shocks: When are 2D simulations enough?} Physics of Plasmas, 32, 052901,
  \dodoi{10.1063/5.0269035}

\bibitem[{D.~V. {Reames}(2013){Reames}}]{reames+13}
{Reames}, D.~V. 2013, \bibinfo{title}{{The Two Sources of Solar Energetic
  Particles},} \ssr, 175, 53, \dodoi{10.1007/s11214-013-9958-9}

\bibitem[{B. {Reville} \& A.~R. {Bell}(2012){Reville} \& {Bell}}]{reville+12}
{Reville}, B., \& {Bell}, A.~R. 2012, \bibinfo{title}{{A filamentation
  instability for streaming cosmic rays},} MNRAS, 419, 2433,
  \dodoi{10.1111/j.1365-2966.2011.19892.x}

\bibitem[{C. Schreiner {et~al.}(2020)Schreiner, Kilian, Spanier, Muñoz, \&
  Büchner}]{schreiner+20}
Schreiner, C., Kilian, P., Spanier, F., Muñoz, P.~A., \& Büchner, J. 2020,
  Ion acceleration in non-relativistic quasi-parallel shocks using fully
  kinetic simulations, \doarXiv{2003.07293}

\bibitem[{V. Tatischeff {et~al.}(2021)Tatischeff, Raymond, Duprat, Gabici, \&
  Recchia}]{tatischeff+21}
Tatischeff, V., Raymond, J.~C., Duprat, J., Gabici, S., \& Recchia, S. 2021,
  \bibinfo{title}{The origin of Galactic cosmic rays as revealed by their
  composition,} Monthly Notices of the Royal Astronomical Society, 508, 1321,
  \dodoi{10.1093/mnras/stab2533}

\bibitem[{K.~J. {Trattner} \& M. {Scholer}(1994){Trattner} \&
  {Scholer}}]{trattner+94}
{Trattner}, K.~J., \& {Scholer}, M. 1994, \bibinfo{title}{{Diffuse minor ions
  upstream of simulated quasi-parallel shocks},} \jgr, 99, 6637,
  \dodoi{10.1029/93JA03165}

\bibitem[{A.~J. {Tylka} {et~al.}(2005){Tylka}, {Cohen}, {Dietrich}, {Lee},
  {Maclennan}, {Mewaldt}, {Ng}, \& {Reames}}]{tylka+05}
{Tylka}, A.~J., {Cohen}, C.~M.~S., {Dietrich}, W.~F., {et~al.} 2005,
  \bibinfo{title}{{Shock Geometry, Seed Populations, and the Origin of Variable
  Elemental Composition at High Energies in Large Gradual Solar Particle
  Events},} \apj, 625, 474, \dodoi{10.1086/429384}

\bibitem[{G. {Zacharegkas} {et~al.}(2024){Zacharegkas}, {Caprioli}, {Haggerty},
  {Gupta}, \& {Schroer}}]{zacharegkas+24}
{Zacharegkas}, G., {Caprioli}, D., {Haggerty}, C., {Gupta}, S., \& {Schroer},
  B. 2024, \bibinfo{title}{{Modeling the Saturation of the Bell Instability
  Using Hybrid Simulations},} \apj, 967, 71, \dodoi{10.3847/1538-4357/ad3960}

\bibitem[{V.~N. {Zirakashvili} \& V.~S. {Ptuskin}(2008){Zirakashvili} \&
  {Ptuskin}}]{zirakashvili+08}
{Zirakashvili}, V.~N., \& {Ptuskin}, V.~S. 2008, \bibinfo{title}{{Diffusive
  Shock Acceleration with Magnetic Amplification by Nonresonant Streaming
  Instability in Supernova Remnants},} ApJ, 678, 939, \dodoi{10.1086/529580}

\end{thebibliography}
\bibliographystyle{aasjournalv7}

\end{document}